\documentstyle[twoside]{article}

\catcode`\@=11
\long\def\@makefntext#1{
\protect\noindent \hbox to 3.2pt {\hskip-.9pt  
$^{{\eightrm\@thefnmark}}$\hfil}#1\hfill}		%CAN BE USED 

\def\@makefnmark{\hbox to 0pt{$^{\@thefnmark}$\hss}}	%ORIGINAL 
	
\def\ps@myheadings{\let\@mkboth\@gobbletwo
\def\@oddhead{\hbox{}
\rightmark\hfil\eightrm\thepage}   
\def\@oddfoot{}\def\@evenhead{\eightrm\thepage\hfil
\leftmark\hbox{}}\def\@evenfoot{}
\def\sectionmark##1{}\def\subsectionmark##1{}}

\oddsidemargin=\evensidemargin
\newcounter{sectionc}\newcounter{subsectionc}\newcounter{subsubsectionc}
\renewcommand{\section}[1] {\vspace{12pt}\addtocounter{sectionc}{1} 
\setcounter{subsectionc}{0}\setcounter{subsubsectionc}{0}\noindent 
	{\tenbf\thesectionc. #1}\par\vspace{5pt}}
\renewcommand{\subsection}[1] {\vspace{12pt}\addtocounter{subsectionc}{1} 
	\setcounter{subsubsectionc}{0}\noindent 
	{\bf\thesectionc.\thesubsectionc. {\kern1pt \bfit #1}}\par\vspace{5pt}}
\renewcommand{\subsubsection}[1] {\vspace{12pt}\addtocounter{subsubsectionc}{1}
	\noindent{\tenrm\thesectionc.\thesubsectionc.\thesubsubsectionc.
	{\kern1pt \tenit #1}}\par\vspace{5pt}}
\newcommand{\nonumsection}[1] {\vspace{12pt}\noindent{\tenbf #1}
	\par\vspace{5pt}}

\newcommand{\textlineskip}{\baselineskip=13pt}
\newcommand{\smalllineskip}{\baselineskip=10pt}

\def\eightcirc{
\begin{picture}(0,0)
\put(4.4,1.8){\circle{6.5}}
\end{picture}}
\def\eightcopyright{\eightcirc\kern2.7pt\hbox{\eightrm c}} 

\newcommand{\copyrightheading}[1]
	{\vspace*{-2.5cm}\smalllineskip{\flushleft
        {\footnotesize Los Alamos electronic archives: gr-qc/9705068r #1}\\
        {\footnotesize $\eightcopyright$\, H.C. Rosu (1998);
        Mod. Phys. Lett. A 13 (1998) 227-230
        }\\
	 }}

\newcommand{\publisher}[2]{{\begin{center}\footnotesize\smalllineskip 
	Received #1\\
	Revised #2
	\end{center}
	}}

\def\abstracts#1#2#3{{
	\centering{\begin{minipage}{4.5in}\baselineskip=10pt\footnotesize
	\parindent=0pt #1\par 
	\parindent=15pt #2\par
	\parindent=15pt #3
	\end{minipage}}\par}} 

\newcommand{\bibit}{\nineit}

\renewenvironment{thebibliography}[1]
	{\frenchspacing
	 \ninerm\baselineskip=11pt
	 \begin{list}{\arabic{enumi}.}
        {\usecounter{enumi}\setlength{\parsep}{0pt}     
	 \setlength{\leftmargin 12.7pt}{\rightmargin 0pt} %FOR 1--9 ITEMS
         \setlength{\itemsep}{0pt} \settowidth
	{\labelwidth}{#1.}\sloppy}}{\end{list}}

\newcounter{itemlistc}
\newcounter{romanlistc}
\newcounter{alphlistc}
\newcounter{arabiclistc}

\def\@citex[#1]#2{\if@filesw\immediate\write\@auxout
	{\string\citation{#2}}\fi
\def\@citea{}\@cite{\@for\@citeb:=#2\do
	{\@citea\def\@citea{,}\@ifundefined
	{b@\@citeb}{{\bf ?}\@warning
	{Citation `\@citeb' on page \thepage \space undefined}}
	{\csname b@\@citeb\endcsname}}}{#1}}

\newif\if@cghi
\def\cite{\@cghitrue\@ifnextchar [{\@tempswatrue
	\@citex}{\@tempswafalse\@citex[]}}
\def\citelow{\@cghifalse\@ifnextchar [{\@tempswatrue
	\@citex}{\@tempswafalse\@citex[]}}
\def\@cite#1#2{{$\null^{#1}$\if@tempswa\typeout
	{IJCGA warning: optional citation argument 
	ignored: `#2'} \fi}}

\def\@refcitex[#1]#2{\if@filesw\immediate\write\@auxout
	{\string\citation{#2}}\fi
\def\@citea{}\@refcite{\@for\@citeb:=#2\do
	{\@citea\def\@citea{, }\@ifundefined
	{b@\@citeb}{{\bf ?}\@warning
	{Citation `\@citeb' on page \thepage \space undefined}}
	\hbox{\csname b@\@citeb\endcsname}}}{#1}}

\def\@refcite#1#2{{#1\if@tempswa\typeout
        {IJCGA warning: optional citation argument
	ignored: `#2'} \fi}}

\def\refcite{\@ifnextchar[{\@tempswatrue
	\@refcitex}{\@tempswafalse\@refcitex[]}}

\def\pmb#1{\setbox0=\hbox{#1}
	\kern-.025em\copy0\kern-\wd0
	\kern.05em\copy0\kern-\wd0
	\kern-.025em\raise.0433em\box0}

\def\fnt#1#2{\footnotetext{\kern-.3em
	{$^{\mbox{\scriptsize #1}}$}{#2}}}

\font\tenrm=cmr10
\font\tenit=cmti10 
\font\tenbf=cmbx10
\font\bfit=cmbxti10 at 10pt
\font\ninerm=cmr9
\font\nineit=cmti9

\font\eightrm=cmr8

\def\qed{\hbox{${\vcenter{\vbox{			%HOLLOW SQUARE
   \hrule height 0.4pt\hbox{\vrule width 0.4pt height 6pt
   \kern5pt\vrule width 0.4pt}\hrule height 0.4pt}}}$}}

\begin{document}

%\runninghead{H.C. Rosu
%$\ldots$} {H.C. Rosu
%$\ldots$}

%Comment (HCR): produce fraza de mai sus la inceputul fiecarei pagini

\normalsize\textlineskip
\thispagestyle{empty}
\setcounter{page}{1}

\copyrightheading{}			%{Vol. 0, No.0 (1992) 000--000}

\vspace*{0.88truein}

%\fpage{1} %%%%%%%%%%%%%%%%%%%%%%%%%%%%%%%%%%%%%%%%%%%%%%%%%%%%%%%%%%%
\centerline{\bf DARBOUX PARAMETER
     FOR EMPTY FRW QUANTUM UNIVERSES}
\centerline{\bf  AND QUANTUM COSMOLOGICAL SINGULARITIES}
\vspace*{0.035truein}
%\centerline{\bf MANUSCRIPTS USING COMPUTER SOFTWARE\footnote{For
%the title, try not to use more than 3 lines. Typeset the title
%in 10 pt Times Roman, uppercase and boldface.}}
\vspace*{0.37truein}
\centerline{\footnotesize HARET C. ROSU}
%\footnote{Typeset names in
%10 pt Times Roman, uppercase. Use the footnote to indicate the
%present or permanent address of the author.}}
\vspace*{0.015truein}
\centerline{\footnotesize\it Instituto de F\'{\i}sica,
Universidad de Guanajuato, Apdo Postal E-143, Le\'on, Gto, Mexico}
\baselineskip=10pt
%\centerline{\footnotesize\it City, State ZIP/Zone,
%Country\footnote{State completely without abbreviations, the
%affiliation and mailing address, including country. Typeset in 8
%pt Times Italic.}}
\vspace*{10pt}
%\centerline{\footnotesize SECOND AUTHOR}
%\vspace*{0.015truein}
%\centerline{\footnotesize\it Group, Laboratory, Address}
%\baselineskip=10pt
%\centerline{\footnotesize\it City, State ZIP/Zone, Country}
\vspace*{0.225truein}
\publisher{(May 27, 1997)}{(January 17, 1998)}

\vspace*{0.21truein}
\abstracts{I present the factorization(s) of the Wheeler-DeWitt equation
for vacuum FRW minisuperspace universes of arbitrary Hartle-Hawking factor
ordering, including the so-called strictly isospectral supersymmetric
method. By the latter means, one can introduce an infinite class of singular
FRW minisuperspace
wavefunctions characterized by a Darboux parameter
that mathematically speaking is a Riccati integration
constant, while physically determines the position of these strictly
isospectral singularities on the Misner time axis.
}{}{}

%\vspace*{10pt}
%\keywords{The contents of the keywords}

\textlineskip                  %) USE THIS MEASUREMENT WHEN THERE IS
\vspace*{12pt}                 %) NO SECTION HEADING

\vspace*{1pt}\textlineskip	%) USE THIS MEASUREMENT WHEN THERE IS
%\section{General Appearance}    %) A SECTION HEADING
\vspace*{-0.5pt}
\noindent

%%%%%%%%%%%%%%%%%%%%%%%%%%%%%%%%%%%%%%%%%%%%%%
%PACS number(s):  98.80.Hw, 11.30.Pb

\noindent
%%%%%%%%%%%%%%%%%%%%%%%%%%%%%%%%%%%%%%%%%%%%%%%%%%%%%%%%%%%%%%%%%%%%%

%\newpage

%\pagebreak

%\textheight=7.8truein
%\setcounter{footnote}{0}
%\renewcommand{\thefootnote}{\alph{footnote}}

%\section{The Main Text}
\noindent

In cosmology one can encounter arbitrary parameters that are difficult
to determine by experiment and theory.
One famous example is Einstein's cosmological constant, which does set the
time scale in the vacuum dominated universe.$^{1}$
Another, very recent example is Immirzi's parameter in quantum general
relativity.$^{2}$
In this work, using the factorization(s) of the Wheeler-DeWitt (WDW)
minisuperspace equation for the vacuum
Friedmann-Robertson-Walker (FRW) universes,
an infinite class of singular FRW wavefunctions are introduced
containing an
arbitrary parameter of Darboux type$^{3}$ determining the position of the
singularities during their Misner time evolution.

Recalling that factorizations of second order linear one-dimensional
differential operators are common
tools in Witten's supersymmetric quantum mechanics$^{4}$ and imply
particular solutions of Riccati equations known as superpotentials,
I explore here the physical consequences of the most general factoring
of the FRW WDW equation
for arbitrary Hartle-Hawking$^{5}$ factor ordering $Q$
by means of the general solution of Riccati equation, a
procedure which has been first used in physics by Mielnik %\cite{M}
for the
quantum harmonic oscillator,$^{6}$ and which is known as the strictly
isospectral supersymmetric method and/or the double Darboux method.$^{7}$
Stated differently, I shall exploit the
non-uniqueness of the factorization of second-order linear differential
operators on the simple example of the FRW WDW differential equation with
Hartle-Hawking factor ordering, that I write down in the form
%%%%%%%%%%%%%%
\begin{equation}
D^{2}\Psi\equiv
\left(\frac{d^2}{d\Omega^2}+2\frac{Q}{2}\frac{d}{d\Omega}-
e^{-4\Omega}   %+(Q^2/4)
\right)\Psi= 0~.
\label{1}
\end{equation}
%%%%%%%%%%%%%%%%
The independent variable $\Omega$ is Misner's time
parameter related to the volume of space $V$ at a given cosmological
epoch through $\Omega =-\ln (V^{1/3})$.$^{8}$
Using the change of function $\Psi=\Phi \exp(-\frac{Q}{2}\Omega)$, one gets the
following Schr\"odinger equation of zero eigenvalue on the Misner axis
%%%%%%%%%%
\begin{equation}
\Phi ^{"}-\left(\frac{Q^2}{4}+e^{-4\Omega}\right)\Phi=0~.
\label{2}
\end{equation}
%%%%%%%%%%%%%%%
The general solution is the modified Bessel function
$Z_{Q/4}(i\frac{1}{2}e^{-2\Omega})= C_1 I_{Q/4}(\frac{1}{2}e^{-2\Omega})+
C_2K_{Q/4}(\frac{1}{2}e^{-2\Omega})$ (the Cs are superposition constants).
%$I_0$ and $K_0$ are the usual modified Bessel functions)
%that later we shall consider in its particular forms.
Usually, the strictly isospectral supersymmetric technique
requires nodeless, normalizable Schr\"odinger solutions.
Although the modified Bessel functions $Z_{Q/4}$ are not normalizable, one can
still use the strictly isospectral technique as first shown
by Pappademos {\em et al} for a few simple scattering
examples in ordinary quantum mechanics.$^{9}$
%and try to implement in
%quantum cosmology the concept of bound states embedded at zero energy,
%which may be considered as
%a more restrictive case of the bound states in the continuum introduced
%by von Neumann and Wigner in nonrelativistic quantum mechanics as early as
%1929 \cite{nw}.
Using the modified Bessel function $Z_{Q/4}$, the
strictly isospectral supersymmetric construction says that the class of
strictly isospectral potentials for
the FRW cosmological models are given by
%%%%%%%%%%%%%%%%%%
\begin{eqnarray} \nonumber
S_{iso}(\Omega;\Omega _{D}) &=&
S(\Omega)-2[\ln({\cal J} _{Z}(\Omega)+\Omega _{D})]''\\ \label{s2}
&=&
S(\Omega)-\frac{4Z_{Q/4}Z_{Q/4}^{'}}{{\cal J} _{Z}+\Omega _{D}}+
\frac{2Z_{Q/4}^4}{({\cal J} _{Z}+\Omega _{D})^2}~,
\end{eqnarray}
where $S(\Omega)=\frac{Q^2}{4}+e^{-4\Omega}$, $\Omega _{D}$
is the Darboux parameter introduced by this method and
%%%%%%%%%%%%%%%%%%
\begin{equation} \label{11}
{\cal J} _{Z}(\Omega)\equiv\int_{\infty}^{\Omega}
Z_{Q/4}^2(ie^{-2y}/2)dy~.
\end{equation}
%%%%%%%%%%%%%%%%%%%%
To get (3) one factorizes the one-dimensional FRW
Schr\"odinger equation with the operators
 $A=\frac{d}{d\Omega}+{\cal W}(\Omega)$ and
$A^{\dagger}=-\frac{d}{d\Omega}+{\cal W}(\Omega)$,
 where the superpotential function
is given by ${\cal W}=-Z_{Q/4}^{'}/Z_{Q/4}$. The Schr\"odinger
potential
and Witten's superpotential enter an initial `bosonic' Riccati equation
$S={\cal W} ^2-{\cal W} ^{'}$. On the other hand, one can build a `fermionic'
Riccati equation $S^{+}={\cal W} ^2+{\cal W} ^{'}$,
 corresponding to a
`fermionic' Schr\"odinger equation for which the operators
 $A$ and $A^{+}$
are applied in reversed order.
 Thus, the `fermionic' FRW Schr\"odinger potential
is found to be
 $S^{+}=\frac{Q^2}{4}+e^{-4\Omega}-2 (\frac{Z_{Q/4}^{'}}{Z_{Q/4}})^{'}$.
This potential does not have
%the  %particular
$Z_{Q/4}$ as an eigenfunction. %solution as a `wavefunction of the universe'.
 However, it is possible
to reintroduce the $Z_{Q/4}$ solution into the spectrum, by
using the general superpotential solution of the fermionic Riccati
equation. The general Riccati solution reads
%%%%%%%%%%
\begin{equation} \label{12}
{\cal W} _{gen}={\cal W}(\Omega) +
 \frac{d}{d\Omega}\ln [{\cal J} _{Z}(\Omega)+\Omega _{D}]~,
\end{equation}
%%%%%%%%%%%
where $\Omega _{D}\in (-\infty , +\infty)$
occurs as an integration constant.$^{10}$
The way to obtain (5) is well known since the work of Mielnik
and will not be repeated here.
From it one can easily get (3). The Darboux parameter family of FRW
wavefunctions differs from the initial one, being the
$\Omega _{D}$-dependent quotient$^{11}$
 %%%%%%%%%%%%%
\begin{equation} \label{13}
Z_{gen}(\Omega;\Omega _{D})=
\frac{Z_{Q/4}(i\frac{1}{2}e^{-2\Omega})}{{\cal J} _{Z}+\Omega _{D}}~.
\end{equation}
%%%%%%%%%%%%%%%
The last equation has interesting implications for the
celebrated issue of cosmological singularities.
Indeed, the strictly isospectral supersymmetric method may be considered as
a way of generating parametric families of new cosmological wavefunctions
with the important feature that there is always a point $\Omega _{s}$ on
the Misner axis defined by $\Omega _{D}=-{\cal J} _{Z} (\Omega _{s})$
at which
they become singular (as well as the strictly isospectral potentials).
In other words, an infinite set of such
strictly isospectral supersymmetric singularities occur, whereas in the
classical case
one is usually bound only to the idea of the original singularity, which is
by definition at $\Omega \rightarrow \infty$, ``an infinite $\Omega$ time in
the past" as Misner put it.$^{8}$
Therefore, we agree with Wheeler's dictum$^{12}$
``No, quantum mechanics does not
provide an escape from gravitational collapse". Even more, the solutions
with singularities are never a set of measure zero for the quantum FRW
universes. There exist well-defined mathematical procedures to produce
an infinite amount of solutions with singularities, at least for the FRW case.
Finally, for some plots displaying the strictly isospectral supersymmetric
FRW singularities in the particular case $Q=0$
the reader is directed to one of my works with Socorro.$^{13}$

\nonumsection{Acknowledgements}
\noindent
%\section*{Acknowledgment}
This work was partially supported by the CONACyT Project 4868-E9406.
The author thanks the referee for suggestions and Dr. B.K. Berger
for a useful electronic message.

%This section should come before the References. Funding
%information may also be included here.

\nonumsection{References}
%\noindent
%References are to be listed in the order cited in the text. Use
%the style shown in the following examples. For journal names,
%use the standard abbreviations. Typeset references in 9 pt Times
%Roman.

%\appendix

%\noindent
%Appendices should be used only when absolutely necessary. They
%should come after the References. If there is more than one
%appendix, number them alphabetically. Number displayed equations
%occurring in the Appendix in this way, e.g.~(\ref{that}), (A.2),
%etc.
%\begin{equation}
%\mu(n, t) = {\sum^\infty_{i=1} 1(d_i < t, N(d_i) = n) \over
%\int^t_{\sigma=0} 1(N(\sigma) = n)d\sigma}\,. \label{that}
%\end{equation}
\end{document}